\documentstyle[12pt]{article}

\newcommand{\be}{\begin{equation}}
\newcommand{\ee}{\end{equation}}
\newcommand{\bea}{\begin{eqnarray}}
\newcommand{\eea}{\end{eqnarray}}

\def\thebibliography#1{\centerline{\bf References}
\list
 {[\arabic{enumi}]}{\settowidth\labelwidth{[#1]}\leftmargin\labelwidth
 \advance\leftmargin\labelsep
 \usecounter{enumi}}
 \def\newblock{\hskip .11em plus .33em minus .07em}
 \sloppy\clubpenalty4000\widowpenalty4000
 \sfcode`\.=1000\relax}

\newcommand{\bi}[1]{\vspace{-2mm} \bibitem{#1}}

\begin{document}
 \thispagestyle{empty}
 \begin{flushright}
% {MZ-TH--95-13}\\[5mm]
 {\tt University of Bergen, Department of Physics}    \\[2mm]
 {\tt Scientific/Technical Report No.1995-06}    \\[2mm]
 {\tt ISSN 0803-2696} \\[5mm]
% {BERGEN--1995-06} \\[5mm]
 {MZ-TH--95-13}\\[5mm]
 {hep-ph/9504432} \\[5mm]
 {April 1995}           \\
\end{flushright}
 \vspace*{3cm}
 \begin{center}
 {\bf \large
Two-loop vacuum diagrams and tensor decomposition
}\footnote{To appear in Proceedings of the AIHENP-95 conference,
Pisa, April 1995 (World Scientific, Singapore,\\
$\hspace*{6mm}$ 1995)}
 \end{center}
 \vspace{1cm}
 \begin{center}
 A.~I.~Davydychev$^{a,}$\footnote{On leave from
 Institute for Nuclear Physics,
 Moscow State University, 119899, Moscow, Russia.\\
 $\hspace*{6mm}$ E-mail address: davyd@vsfys1.fi.uib.no}
 \ \ and \ \
J.~B.~Tausk$^{b,}$\footnote{E-mail address:
tausk@vipmzw.physik.uni-mainz.de}\\
 \vspace{1cm}
$^{a}${\em
 Department of Physics, University of Bergen, \\
      All\'{e}gaten 55, N-5007 Bergen, Norway}
\\
\vspace{.3cm}
$^{b}${\em
Institut f\"ur Physik,
  Johannes Gutenberg Universit\"at, \\
Staudinger Weg 7, D-55099 Mainz, Germany}
\end{center}
 \hspace{3in}
 \begin{abstract}
General algorithms for tensor reduction of two-loop massive
vacuum diagrams are discussed. Some explicit useful formulae
are presented.
 \end{abstract}

\newpage

\vspace*{5mm}

{\bf 1. General Remarks}

\vspace{3mm}

In the talk given by one of the authors (A.~D.) at the AIHENP-95
Conference in Pisa, some recent (i.e. since AIHENP-93) developments
in two-loop calculations were reviewed.
The main results discussed in the talk were
published \cite{UD3-4,BDST,Magic} and were obtained in collaboration
with N.I.~Ussyukina \cite{UD3-4}, F.A.~Berends and
V.A.~Smirnov \cite{BDST}, and the other author of this
paper, J.B.~T.  \cite{BDST,Magic}.

In ref.~\cite{UD3-4} (see also in ref.~\cite{UD1-2}), some new results
for two-loop three-point diagrams with massless internal particles
were obtained, including the diagrams with irreducible numerators.
The paper \cite{BDST} continues the project started in refs. \cite{DT,DST}
and provides explicit results for the small momentum expansion
of two-loop self-energy diagrams for all cases when the smallest
physical threshold is zero (``zero-threshold expansion'').
To do this, general results on asymptotic expansions of Feynman
diagrams were applied (see e.g. in the reviews \cite{asex} and
references therein). In the paper \cite{Magic}, we have established
a useful connection between two-loop massive vacuum diagrams and
off-shell massless triangle diagrams. This connection is valid for
all values of the space-time dimension $n$ and masses $m_i$ (which are
related to the external momenta squared of the triangle diagram
via $m_i^2=p_i^2$, $i=1,2,3$). As a corollary, the analytic result
for the $\varepsilon$-part ($\varepsilon\equiv (4-n)/2$) of
two-loop vacuum diagram was obtained in terms of trilogarithms.
For the case of equal masses, a new transcendental constant
(related to log-sine integral $\mbox{Ls}_3(\textstyle{2\over3}\pi)$)
is shown to appear.

Since it is impossible to give a detailed description of all these
results in a short contribution, and also because the main results
have been already published, we decided to concentrate ourselves on
an important particular problem related to two-loop vacuum diagrams,
namely the problem of tensor decomposition.

Tensor reduction of Feynman integrals containing loop momenta with
uncontracted Lorentz indices in the numerator is very important
for various realistic calculations in the Standard Model (and beyond).
For one-loop diagrams with different numbers of external lines,
several approaches and algorithms were developed \cite{one-tensor}.
For two-loop vacuum and self-energy diagrams, the problem
was considered e.g. in refs. \cite{two-tensor,Ch-AI93}, whilst the
three-point two-loop case is more complicated and requires
results for the integrals with irreducible numerators \cite{UD3-4}.

The problem of finding general algorithms for the tensor decomposition of
two-loop vacuum diagrams is interesting because it is connected with the
calculation of coefficients of the small momentum expansion \cite{DT},
also for the three-point case \cite{FT}. In the three-point case, the
problem is more tricky (even for scalar integrals), because we
have two independent external momenta to contract with. Some relevant
formulae for cases when the numerator is contracted with one or two
external vectors can be found in refs. \cite{DST,FT2}. We also note that
the general result for a special case (when two masses are equal and the third
is zero) was presented in ref.~\cite{Ch-AI93}.

%\pagebreak

\newpage
%\vspace{5mm}

{\bf 2. Tensor Decomposition}

\vspace{3mm}

In this paper, we shall use the following notation:
\begin{equation}
\label{defI}
I\left[ \mbox{something} \right]
\equiv
\int\int \mbox{d}^n p \; \mbox{d}^n q \;
\left\{ \mbox{something} \right\} \;
F\left(p^2, q^2, (pq)\right) ,
\end{equation}
and we are interested in expressing the integrals
\begin{equation}
\label{I}
I\left[ p_{\mu_1} \ldots p_{\mu_{N_1}} \;
         q_{\sigma_1} \ldots q_{\sigma_{N_2}} \right]
\end{equation}
in terms of scalar integrals. In eq.~(\ref{defI}),
$F\left(p^2, q^2, (pq)\right)$ is an arbitrary scalar function
depending on Lorentz invariants of the loop momenta $p$ and $q$.
In the special case when this function does not depend on $(pq)$
we shall write it as $F\left(p^2, q^2\right)$.
Usually, this function is a product of propagators,
\begin{equation}
\label{denom}
\left( p^2\!-\!m_1^2 \right)^{-\nu_1}
\left( q^2\!-\!m_2^2 \right)^{-\nu_2}
\left( (p\!-\!q)^2\!-\!m_3^2 \right)^{-\nu_3} \; \;
\mbox{or} \; \;
\left( p^2\!-\!m_1^2 \right)^{-\nu_1}
\left( q^2\!-\!m_2^2 \right)^{-\nu_2} \; ,
\end{equation}
but all formulae we are going to discuss are valid for
arbitrary scalar functions.

Some explicit results for the cases when the integral (\ref{I})
was contracted with the external vector (momentum) $k$
were given in ref.~\cite{DST} (eqs.~(B.10) and (B.9)).
They are also valid when the integrands (\ref{denom}) are
replaced by $F\left(p^2, q^2, (pq)\right)$ (in eq.~(B.10))
or $F\left(p^2, q^2\right)$ (in eq.~(B.9)), respectively.
Moreover, we note that eq.~(B.9) can be generalized to the
case with two external momenta $k_1$ and $k_2$,
\begin{eqnarray}
\label{B.9+}
\left. \int \int \mbox{d}^n p \; \mbox{d}^n q
\; F\left(p^2, q^2\right)
\; [2(k_1 p)]^{N_1} \; [2(k_2 q)]^{N_2} \; [2(p q)]^{N_3}
\right|_{ \begin{array}{c}
    {}_{N_1+N_3  \mbox{-- \small even}} \\[-1mm]
    {}_{N_2+N_3  \mbox{-- \small even}}
         \end{array} }
\hspace{1cm} \nonumber \\
= \frac{N_1 !\; N_2 !\; N_3 !}
{{(n/2)}_{(N_1+N_2)/2}\;{(n/2)}_{(N_1+N_3)/2}\;{(n/2)}_{(N_2+N_3)/2}\;}
\hspace{30mm} \nonumber \\[-3mm]
\times
\begin{array}{c} {} \\ {} \\ \sum \\ {}_{2j_1+j_3=N_1} \\
                {}_{2j_2+j_3=N_2}  \end{array}
\frac{{(k_1^2)}^{j_1} {(k_2^2)}^{j_2} {[2(k_1 k_2)]}^{j_3}}
     {j_1 ! \; j_2 ! \; j_3 !} \;
\frac{{(n/2)}_{(N_1+N_2+N_3-j_3)/2}}{((N_3-j_3)/2)!}
\nonumber \\
 \times \int \int \mbox{d}^n p \; \mbox{d}^n q
\; F\left(p^2, q^2\right)
\; {(p^2)}^{(N_1 + N_3)/2} \; {(q^2)}^{(N_2 + N_3)/2} \; ,
\end{eqnarray}
where $(a)_j \equiv \Gamma(a+j)/\Gamma(a)$ denotes the
Pochhammer symbol.
If $(N_1+N_3)$ or $(N_2+N_3)$ is odd,
the integral on the l.h.s. is equal to zero.

Since we have no external momenta in (\ref{I}),
only tensor structures constructed of metric tensors may be
involved in the tensor decomposition. Therefore, (\ref{I}) vanishes
if $N\equiv N_1+N_2$ is odd. For even $N$, the suitable tensor
structures should be symmetric with respect to two subsets
of indices, $(\mu_1, \ldots , \mu_{N_1})$ and
$(\sigma_1, \ldots , \sigma_{N_2} )$.
All independent tensor structures can be constructed in the
following way. Let us take a product of $j_1$ metric tensors
$g_{\mu_i \mu_k}$, $j_2$ tensors $g_{\sigma_i \sigma_k}$
and $j_3$ tensors $g_{\mu_i \sigma_k}$
(we should remember that the conditions $2 j_1+j_3=N_1$ and
$2 j_2+j_3=N_2$ should be satisfied).
Now, let us consider all possible permutations of $\mu$'s
and all possible permutations of $\sigma$'s producing
{\em distinct} products of metric tensors,
%(except for interchanging the indices
%belonging to the same metric tensor),
and take the sum of all these terms (with the coefficients equal to one).
So, the obtained tensor structure is symmetrized in
$\mu$'s and $\sigma$'s, and it is
\begin{eqnarray}
\label{j1j2j3}
%{} \nonumber \\[-8mm]
\! \left\{ j_1, j_2, j_3 \right\}
\equiv
g_{\mu_1 \mu_2} \ldots g_{\mu_{2j_1\!-\!1} \mu_{2j_1}}
g_{\sigma_1 \sigma_2} \ldots g_{\sigma_{2j_2\!-\!1} \sigma_{2j_2}}
g_{\mu_{2j_1\!+\!1} \sigma_{2j_2\!+\!1}} \ldots
g_{\mu_{2j_1\!+\!j_3} \sigma_{2j_2\!+\!j_3}}
\nonumber \\
+ \mbox{permutations} .
\end{eqnarray}

For given $N_1$ and $N_2$, the number of
independent tensor structures (\ref{j1j2j3}) is
\begin{equation}
\label{T}
T(N_1, N_2) = \min\left( \left[\textstyle{1\over2}N_1\right],
\left[\textstyle{1\over2}N_2\right] \right) +1 ,
\end{equation}
where $\left[\textstyle{1\over2}N_i\right]$ is the
integer part of $N_i/2$,
and the number of terms on the r.h.s. of (\ref{j1j2j3}) is
\begin{equation}
\label{c}
%{} \nonumber \\[-9mm]
c_{j_1 j_2 j_3} = \frac{{N_1}! \; {N_2}!}
                       {2^{j_1+j_2} \; {j_1}! \; {j_2}! \; {j_3}!} .
\end{equation}
Note that due to the conditions $2 j_1+j_3=N_1$ and $2 j_2+j_3=N_2$,
at given $N_1$ and $N_2$ the tensor structures (\ref{j1j2j3})
are completely defined by one index, $j_3$.

So, the result for the integral (\ref{I})
(for even $N_1+N_2$) should look like
\begin{eqnarray}
\label{decomp1}
{} \nonumber \\[-8mm]
I\left[ p_{\mu_1} \ldots p_{\mu_{N_1}} \;
         q_{\sigma_1} \ldots q_{\sigma_{N_2}} \right]
= \begin{array}{c} {} \\ {} \\ \sum \\ {}_{2j_1+j_3=N_1} \\
                {}_{2j_2+j_3=N_2}  \end{array}
\left\{ j_1, j_2, j_3 \right\} \; I_{j_1 j_2 j_3} \; ,
\end {eqnarray}
where $I_{j_1 j_2 j_3}$ are some scalar integrals which are to be found.
A standard way to define $I_{j_1 j_2 j_3}$ is to consider all
independent contractions of (\ref{decomp1}) (e.g. to contract it with
each of the structures (\ref{j1j2j3})). In such a way, one needs to
solve a system of $T(N_1,N_2)$ (see eq.~(\ref{T})) linear equations.

Due to power-counting reasons, the scalar integrals $I_{j_1 j_2 j_3}$
should be linear combinations of the integrals
(\ref{defI}) with scalar numerators, carrying the same
total powers of the loop momenta $p$ and $q$. Therefore,
we can re-write eq.~(\ref{decomp1}) as
\begin{eqnarray}
\label{decomp2}
I\left[ p_{\mu_1} \ldots p_{\mu_{N_1}} \;
         q_{\sigma_1} \ldots q_{\sigma_{N_2}} \right]
\hspace{78mm}
\nonumber \\[-4mm]
= \! \begin{array}{c} {} \\ {} \\ \sum \\ {}_{2j_1+j_3=N_1} \\
                {}_{2j_2+j_3=N_2}  \end{array}
  \left\{ j_1, j_2, j_3 \right\}  \!
\begin{array}{c} {} \\ {} \\ \sum \\ {}_{2j'_1+j'_3=N_1} \\
                {}_{2j'_2+j'_3=N_2}  \end{array}
 \phi_{j_1 j_2 j_3; j'_1 j'_2 j'_3} \; c_{j'_1 j'_2 j'_3} \;
I\left[ (p^2)^{j'_1} (q^2)^{j'_2} (pq)^{j'_3} \right]  ,
\end{eqnarray}
% \begin{eqnarray}
% \label{decomp2}
% I\left[ p_{\mu_1} \ldots p_{\mu_{N_1}} \;
%          q_{\sigma_1} \ldots q_{\sigma_{N_2}} \right]
% %\hspace{50mm}
% %\nonumber \\
% = \frac{1}{2^{(N_1+N_2)/2} \; (n/2)_{(N_1+N_2)/2}} \;
% \hspace{32mm}
% \nonumber \\
% \times
% \begin{array}{c} {} \\ {} \\ \sum \\ {}_{2j_1+j_3=N_1} \\
%                 {}_{2j_2+j_3=N_2}  \end{array}
% c_{j_1 j_2 j_3} \; I\left[ (p^2)^{j_1} (q^2)^{j_2} (pq)^{j_3} \right] \;
% %\nonumber \\
% %\times
% \begin{array}{c} {} \\ {} \\ \sum \\ {}_{2j'_1+j'_3=N_1} \\
%                 {}_{2j'_2+j'_3=N_2}  \end{array}
% f_{j_1 j_2 j_3; j'_1 j'_2 j'_3} \; \left\{ j'_1, j'_2, j'_3 \right\} \; ,
% \end{eqnarray}
% where $(a)_j\equiv \Gamma(a+j)/\Gamma(a)$ is the Pochhammer symbol,
where $c_{j_1 j_2 j_3}$ are defined in (\ref{c}), while
$\phi_{j_1 j_2 j_3; j'_1 j'_2 j'_3}$ can be considered
as the elements of the ``decomposition'' matrix
($T\times T$, see eq.~(\ref{T})).
Contracting eq.~(\ref{decomp2}) with all
possible tensors (\ref{j1j2j3}), we get a ``column''
of $c_{j_1 j_2 j_3} I\left[ (p^2)^{j_1} (q^2)^{j_2} (pq)^{j_3} \right]$
on the l.h.s. On the r.h.s., we get the ``contraction'' matrix
$\chi$ with the elements defined as
\begin{equation}
\label{chi}
\chi_{j_1 j_2 j_3; j'_1 j'_2 j'_3}
= \mbox{contraction}\left( \left\{ j_1, j_2, j_3 \right\} ,
                           \left\{ j'_1, j'_2, j'_3 \right\} \right) .
\end{equation}

When speaking about matrices, we understand that
only one of $j$-indices is
relevant in each set $(j_1,j_2,j_3)$ (e.g., $j_3$).
For given $N_1$ and $N_2$, the values of $j_3$ can be either even
(if $N_1$ and $N_2$ are even) or odd (if $N_1$ and $N_2$ are odd).
In order to use the matrix notation in ordinary form, we introduce
the generalized index $j\equiv (j_1,j_2,j_3)$ (and
$j'\equiv (j'_1,j'_2,j'_3)$, etc.) which takes values from
1 to $T(N_1,N_2)$. In all ``non-matrix'' formulae, however, we would prefer
to keep all the indices $j_1, j_2, j_3$ explicitly.
So, in the matrix notation, a corollary of
eqs.~(\ref{decomp2})--(\ref{chi}) is
\begin{equation}
\label{chiphi}
%{} \nonumber \\[-7mm]
\sum_{j''}^{} \;
\chi_{{j} {j''}} \; \phi_{{j''} {j'}} = \delta_{{j} {j'}} \; .
\end{equation}
In other words, the decomposition matrix $\phi$ is nothing but the
inverse contraction matrix, $\chi^{-1}$. Since $\chi_{{j} {j'}}$
is symmetric (see (\ref{chi})), $\phi_{{j} {j'}}$ is symmetric as well.

Thus, the problem is how to find all $\phi_{j_1 j_2 j_3; j'_1 j'_2 j'_3}$
for given $N_1$ and $N_2$. One way is to use recurrence relations.
%It is convenient to introduce $f_{j_1,j_2,j_3; j'_1,j'_2,j'_3}$
%related to $\chi_{j_1 j_2 j_3; j'_1 j'_2 j'_3}$ via
%\begin{equation}
%f_{j_1,j_2,j_3; j'_1,j'_2,j'_3}
%= 2^{(N_1+N_2)/2} \; (n/2)_{(N_1+N_2)/2} \;
%\phi_{j_1 j_2 j_3; j'_1 j'_2 j'_3} .
%\end{equation}
%Contracting (\ref{decomp2}) with
%$k^{\mu_1} \ldots k^{\mu_{N_1}} k^{\sigma_1} \ldots k^{\sigma_{N_2}}$
%and using eq.~(B.10) of ref.~\cite{DST}, we get
%\begin{eqnarray}
%\label{fc}
%{} \nonumber \\[-12mm]
%\begin{array}{c} {} \\ {} \\ \sum \\ {}_{2j'_1+j'_3=N_1} \\
%                {}_{2j'_2+j'_3=N_2}  \end{array}
%f_{j_1,j_2,j_3; j'_1,j'_2,j'_3} \; c_{j'_1 j'_2 j'_3} = 1 .
%\end{eqnarray}
We note that the following simple formulae of contracting
tensor structures (\ref{j1j2j3}) with respect to two indices
can be derived:
\begin{equation}
\label{contract1}
\! g_{\mu_{N_1} \sigma_{N_2}} \! \left\{ j_1, j_2, j_3 \right\}
= \left( n\!+\!2j_1\!+\!2j_2\!+\!j_3\!-\!1 \right)
  \left\{ j_1, j_2, j_3\!-\!1 \right\} +
  ( j_3\!+\!1)
  \left\{ j_1\!-\!1, j_2\!-\!1, j_3\!+\!1\right\} ,
\! \! \! \!
%\nonumber \\
\end{equation}
\begin{equation}
\label{contract2}
g_{\mu_{N_1\!-\!1} \mu_{N_1}} \left\{ j_1, j_2, j_3 \right\}
= \left( n\!+\!2j_1\!+\!2j_3\!-\!2 \right)
  \left\{ j_1\!-\!1, j_2, j_3 \right\}
  \!+\! 2\left( j_2\!+\!1 \right)
  \left\{ j_1, j_2\!+\!1, j_3\!-\!2 \right\}
\end{equation}
(and also an analogous formula for contraction with
$g_{\sigma_{N_2-1} \sigma_{N_2}}$).
Using these contractions, the following
recurrence relations for $\phi_{j_1 j_2 j_3; j'_1 j'_2 j'_3}$
can be obtained:
\begin{eqnarray}
\label{recur1}
N_1 N_2 \left\{ (n+N_1+N_2-j'_3-1) \;
\phi_{j_1 j_2 j_3; j'_1 j'_2 j'_3}
+ (j'_3-1) \; \phi_{j_1 j_2 j_3; j'_1\!+\!1, j'_2\!+\!1, j'_3\!-\!2} \right\}
\nonumber \\
= j_3 \; \phi_{j_1,j_2,j_3-1; j'_1, j'_2, j'_3-1} \; ,
\end{eqnarray}
\begin{eqnarray}
\label{recur2}
%{} \nonumber \\[-8mm]
N_1 \; (N_1-1)
\left\{ (n+N_1+ j'_3-2) \;\phi_{j_1 j_2 j_3; j'_1 j'_2 j'_3}
+ 2 j'_2 \;\phi_{j_1 j_2 j_3; j'_1-1, j'_2-1,j'_3+2} \right\}
\nonumber \\
= 2\;j_1 \; \phi_{j_1-1,j_2,j_3; j'_1-1,j'_2,j'_3} \; ,
\end{eqnarray}
and also a relation similar to (\ref{recur2}) but with interchanged
indices, $1\leftrightarrow 2$.
In eqs.~(\ref{contract1})--(\ref{recur2}), it is understood that,
whenever any of $j$'s becomes negative, the corresponding tensors
$\left\{j_1,j_2,j_3\right\}$ and $\phi$'s should be taken equal to zero.

If we increase $N_1$ and/or $N_2$ without changing the number of
tensor structures (\ref{T}), it is enough to have only relations
(\ref{recur1})--(\ref{recur2}) to express ``higher'' $\phi$'s in
terms of ``lower'' $\phi$'s. If on this step one extra
tensor structure appears (for example, when we go from
odd $N_1$ and $N_2$ to $(N_1+1)$ and $(N_2+1)$),
we need one more relation between $\phi$'s.
%In this case, the relation (\ref{fc}) can be employed.
This extra relation can be obtained by contracting (\ref{decomp2}) with
$k^{\mu_1} \ldots k^{\mu_{N_1}} k^{\sigma_1} \ldots k^{\sigma_{N_2}}$
and using eq.~(B.10) of ref.~\cite{DST},
\begin{eqnarray}
\label{phi-c}
{} \nonumber \\[-12mm]
\begin{array}{c} {} \\ {} \\ \sum \\ {}_{2j'_1+j'_3=N_1} \\
                {}_{2j'_2+j'_3=N_2}  \end{array}
\phi_{j_1 j_2 j_3; j'_1 j'_2 j'_3} \; c_{j'_1 j'_2 j'_3} =
\frac{1}{2^{(N_1+N_2)/2} \; (n/2)_{(N_1+N_2)/2}} \; .
\end{eqnarray}

The values of the lowest $\phi$'s can be taken from the results
at $N_2=0$ and $N_2=1$,
\begin{equation}
\label{N2=0}
I \left[ p_{\mu_1} \ldots p_{\mu_{N_1}} \right]
= \frac{1}{2^{N_1/2} \; (n/2)_{N_1/2}}
  \left\{ \frac{N_1}{2},0,0 \right\} \;\;
  I \left[ (p^2)^{N_1/2} \right] \; ,
\end{equation}
\begin{equation}
\label{N2=1}
I \left[ p_{\mu_1} \ldots p_{\mu_{N_1}} q_{\sigma_1} \right]
= \frac{1}{2^{(N_1+1)/2} \; (n/2)_{(N_1+1)/2}}
  \left\{ \frac{N_1-1}{2},0,1 \right\}  \;
  I \left[ (p^2)^{(N_1-1)/2} (pq) \right] .
\end{equation}
Using the recurrence procedure, it is also possible to obtain some
less trivial formulae, for $N_2=2$ and $N_2=3$ and arbitrary
$N_1$ (such that $N_1\geq N_2$):
\begin{eqnarray}
\label{N2=2}
I \left[ p_{\mu_1} \ldots p_{\mu_{N_1}} q_{\sigma_1} q_{\sigma_2} \right]
= \frac{1}{2^{(N_1+2)/2} \; (n/2)_{(N_1+2)/2} \; (n-1)}
\hspace{40mm}
\nonumber \\
\times \left\{  \left\{ \frac{N_1}{2},1,0 \right\} \;\;
  I \left[ (n+N_1-1)(p^2)^{N_1/2} q^2 - N_1 (p^2)^{(N_1-2)/2} (pq)^2 \right]
 \right.
\nonumber \\
+ \left. \left\{ \frac{N_1-2}{2},0,2 \right\} \;\;
  I \left[ n (p^2)^{(N_1-2)/2} (pq)^2 - (p^2)^{N_1/2} q^2 \right]
\right\} \; ,
\end{eqnarray}
\begin{eqnarray}
\label{N2=3}
%{} \nonumber \\[-7mm]
I \left[ p_{\mu_1} \ldots p_{\mu_{N_1}}
         q_{\sigma_1} q_{\sigma_2} q_{\sigma_3} \right]
= \frac{1}{2^{(N_1+3)/2} \; (n/2)_{(N_1+3)/2} \; (n-1)}
\hspace{39mm}
\nonumber \\
\! \times \left\{  \left\{ \frac{N_1\!-\!1}{2},1,1 \right\}
  I \left[ (n\!+\!N_1\!-\!2)(p^2)^{(N_1\!-\!1)/2} q^2 (pq)
          - (N_1\!-\!1) (p^2)^{(N_1\!-\!3)/2} (pq)^3 \right]
  \! \right.
\nonumber \\
+ \left. \left\{ \frac{N_1-3}{2},0,3 \right\} \;
  I \left[ (n+2) (p^2)^{(N_1-3)/2} (pq)^3
          - 3 (p^2)^{(N_1-1)/2} q^2 (pq) \right]
\right\} .
\end{eqnarray}

Thus, a recursive procedure of calculating
$\phi_{j_1,j_2,j_3; j'_1 j'_2 j'_3}$ is
constructed which provides a general algorithm to calculate
the tensor integrals (\ref{I}).
It is difficult, however, to generalize
eqs.~(\ref{N2=0})--(\ref{N2=3}) to the case of arbitrary $N_2$.

There is, however, another approach to this problem
which does not involve using recurrence relations.
Let us look again at the decomposition (\ref{decomp2}),
remembering also how the integrals $I$ are defined (\ref{defI}).
Then, let us contract eq.~(\ref{decomp2}) with
$k_1^{\mu_1} \ldots k_1^{\mu_{N_1}} k_2^{\sigma_1}
\ldots k_2^{\sigma_{N_2}}$, and multiply it by
$(k_1 k_2)^{N_3} \; F(k_1^2, k_2^2)$. Now, we can integrate
the resulting expression over $k_1$ and $k_2$
(not over $p$ and $q$!), using the
formula (\ref{B.9+}). As a result of this integration
at different values of $N_3$, we get $T(N_1, N_2)$
independent equations.
Then, considering the scalar integrals
$I\left[(p^2)^{j_1} (q^2)^{j_2} (pq)^{j_3}\right]$
as a basis, we get a set of relations for $\phi$'s.
%Moreover, in these relations the
%resulting scalar integrals over $p$ and $q$ are the same on
%the l.h.s. and on the r.h.s., and therefore can be cancelled.
Finally, taking some linear combinations of these relations,
we get a Kronecker symbol on the r.h.s.
This means that we have found the elements of
the matrix inverse to $\phi_{j j'}$ which is
nothing but the contraction matrix $\chi_{j j'}$
(see eqs.~(\ref{chi})--(\ref{chiphi})).
The obtained expression is
\begin{eqnarray}
\label{chi-result}
\chi_{j_1 j_2 j_3; j'_1 j'_2 j'_3}
= \frac{{N_1}! \; {N_2}!}{{j_1}! {j_2}! {j_3}!}
\sum_{l=0}^{[j'_3/2]}
\frac{(-1)^l \; (j_3\!+\!j'_3\!-2l)! \;((n+N_1+N_2-2l+2)/2)_l }
     {l!\; (j'_3\!-\!2l)! \;((j_3+j'_3-2l)/2)! \;
      (n/2)_{(j_3+j'_3\!-\!2l)/2}}
\nonumber \\
\times
(n/2)_{(N_1+j'_3-2l)/2} \; (n/2)_{(N_2+j'_3-2l)/2} \; .
\end{eqnarray}

Thus, we have obtained an explicit general expression for the inverse of the
decomposition matrix, $(\phi^{-1})_{j j'}$. For the cases of interest,
the matrix (\ref{chi-result}) can be inverted by use of
computer systems for analytical calculations.
Although it would be nicer to get an analogous result for
the ``direct'' matrix $\phi_{j j'}$, the presented approach
can also be considered as a general solution to the problem
of tensor decomposition of two-loop vacuum diagrams.

\vspace{3mm}

{\bf Acknowledgements}

A.~D. is grateful to K.G.~Chetyrkin and O.V.~Tarasov for useful
discussions.
A.~D.'s research and participation in AIHENP-95 conference in Pisa
were supported by the Research Council of Norway.
J.B.T. was supported by the Graduiertenkolleg
``Teilchenphysik'' in Mainz.

\newpage

%\nonumsection{References}
%\noindent

%\vspace{-3mm}

\end{document}